\documentclass[10pt,letterpaper,twocolumn]{article} %% two column, final layout

\usepackage{ol2}
\usepackage[draft,implicit=false]{hyperref}
\usepackage{amsmath}

\begin{document}

\twocolumn[ %% activate for two-column option

\title{2D photonic-crystal  optomechanical nanoresonator}

%% For REVTeX it is possible to automate superscript and e-mail callouts with the superscriptaddress option; see REVTeX4 documentation.

\author{K.~Makles,$^{1}$ T.~Antoni,$^2$, A.~G.~Kuhn$^{1}$, S.~Del\'eglise$^{1,*}$, T.~Briant$^{1}$, P.-F.~Cohadon$^{1}$, R.~Braive$^{3}$, G.~Beaudoin$^{3}$, L.~Pinard$^{4}$, C.~Michel$^{4}$, V.~Dolique$^{4}$, R.~Flaminio$^{4}$, G.~Cagnoli$^{4}$, I.~Robert-Philip$^{3}$, A.~Heidmann$^{1}$}

\address{
$^1$Laboratoire Kastler-Brossel, Sorbonne Universités UPMC, ENS, CNRS,
Collège de France, Campus Jussieu, Paris, France\\
$^2$Ecole Centrale Paris, Laboratoire de  Photonique Quantique et Mol\'eculaire, CNRS \\
$^3$Laboratoire de Photonique et de Nanostructures LPN-CNRS, Marcoussis, France \\
$^4$Laboratoire des Matériaux avanc\'es, CNRS, IN2P3, Villeurbanne, France\\
$^*$Corresponding author: samuel.deleglise@lkb.upmc.fr
}

\begin{abstract}We present the optical optimization of an optomechanical device based on a suspended InP membrane patterned with a 2D near-wavelength grating (NWG) based on a 2D photonic-crystal geometry. We first identify by numerical simulation a set of geometrical parameters providing a reflectivity higher than 99.8\,\% over a 50-nm span. We then study the limitations induced by the finite value of the optical waist and lateral size of the NWG pattern using different numerical approaches. The NWG grating, pierced in a suspended InP 265 nm-thick membrane, is used to form a compact microcavity involving the suspended nano-membrane as end mirror. The resulting cavity has a waist size smaller than 10 $\mu$m and a finesse in the 200 range. It is used to probe the Brownian motion of the mechanical modes of the nanomembrane.\end{abstract}


 ] %% activate for two-column option

\begin{figure*}
%\begin{center}
\centering
\includegraphics[height=6.cm]{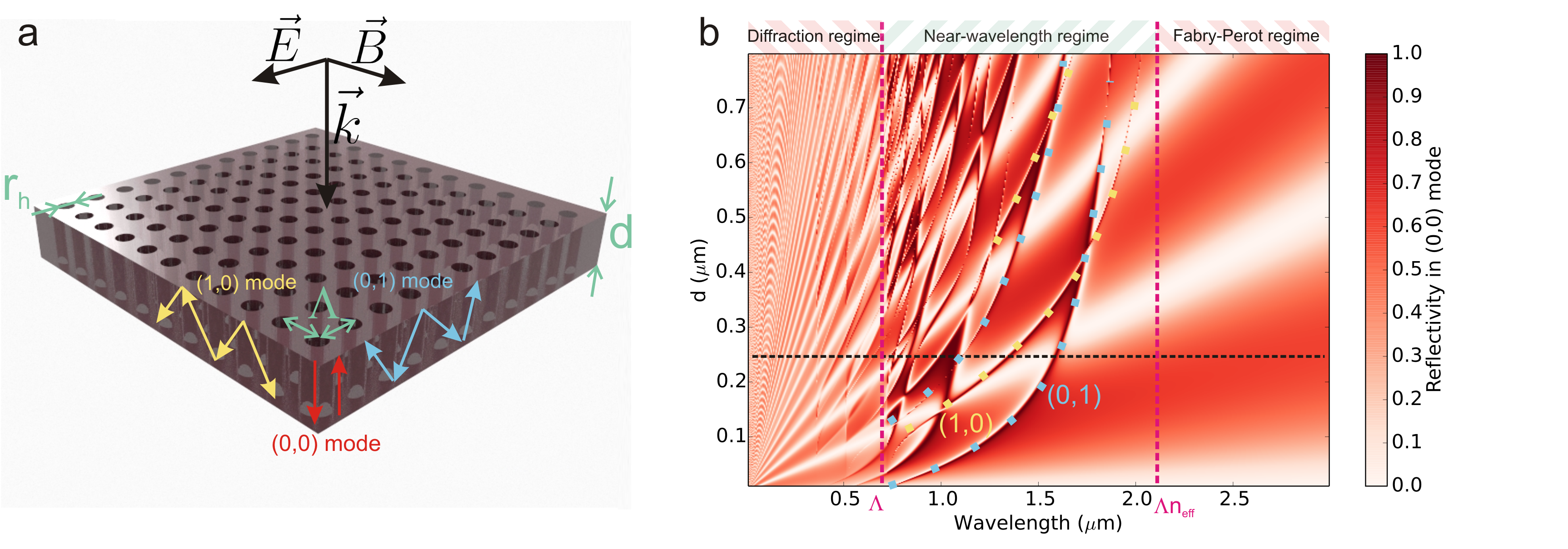}
\caption{ a: Schematic view of the optical modes at play in a 2D near-wavelength grating. b: Reflectivity map at normal incidence obtained by RCWA for an infinite slab with the parameters $r_\mathrm{h}  = 201$ nm, $\Lambda = 725$ nm and $n=3.3$.}
\label{cristal}
%\end{center}
\end{figure*}

\noindent The radiation pressure coupling between optical and mechanical degrees of freedom has been the subject of an intense research effort over the last decade. A large variety of experimental platforms have been developed to observe the quantum effects of radiation pressure, with prospects both in fundamental physics, such as the study of decoherence of macroscopic superpositions \cite{Romero-Isart2011, Pikovski2012}, or in quantum engineering, such as the quantum-coherent conversion of weak signals between different electromagnetic bands \cite{Andrews2014} (e.g. microwave-to-optical  conversion). This program widely relies on the technical challenge of producing mechanical resonators with high optical quality, low mass, and low mechanical dissipation. While traditional free-space optomechanical resonators, based on dielectric materials directly coated on a mechanically-compliant substrate usually suffer from the poor mechanical properties of the $\mathrm{Ta}_2\mathrm{O}_5/\mathrm{SiO}_2$ layers \cite{Gigan2006, Kuhn2011}, in-plane resonators based on semiconductor-derived micro- and nano-fabrication processes \cite{Rokhsari2005, Eichenfield2009} are usually limited by the low input/output coupling efficiencies of the optical cavities \cite{Safavi-Naeini2013}. This constitutes a severe limitation to the efficiency of quantum optical devices. Another successful approach consists in embedding a vibrating dielectric membrane close to the waist of a free-space optical cavity \cite{Thompson2008}. This “membrane-in-the-middle” approach ensures at the same time a good mode matching with the propagating optical beam and uncompromised mechanical properties of the vibrating membrane \cite{Purdy2013, Purdy2013a}. Recent works \cite{Antoni2011, Kemiktarak2012, Kemiktarak2012a} have exploited a patterning of the vibrating membrane by periodic wavelength-scale structures in order to dramatically increase its optical reflectivity. This is beneficial as the optomechanical coupling scales linearly with the reflectivity of the membrane \cite{Jayich2008}. Moreover, the reflectivity plays a crucial role in experiments aiming at the observation of mechanical quantum jumps~\cite{Jayich2008}. 

Here, we present the optical optimization of a 2D-patterned micrometer-scale suspended membrane pierced by a 2D photonic crystal acting as a NWG. Thanks to a specifically designed input mirror with a radius of curvature on the 100-$\mu$m range, we have been able to form a microcavity using the suspended membrane as end mirror and to monitor the Brownian motion of the membrane. %This two-mirror setup constitutes a first step towards integrated, micron-scale membrane-in-the-middle setups allowing to combine the advantages of free-space optomechanical setups with the integrability of micro-fabricated optomechanical structures.

Near-wavelength diffraction gratings \cite{Karagodsky2012} are thin slabs of material whose refractive index is patterned with a period $\Lambda$ slightly smaller than the incident optical wavelength $\lambda$. The diffraction properties of such optical elements are radically different from those of usual diffraction gratings. These properties result from a combination of physical circumstances: $\Lambda<\lambda$ implies that all non-zero order diffracted beams are evanescent in the surrounding air. On the other hand, when $\lambda/n_\text{eff}<\Lambda$  ($n_\text{eff}$ is an effective refractive index of the patterned slab), several propagating modes exist inside the slab and for some particular values of the slab thickness, the boundary conditions at the slab/air interfaces give rise to a destructive interference for the transmitted 0-order beam. By conservation of the optical flux, such a phenomenon is associated with a close to unity reflectivity for the incoming wave. Apart from their application to optomechanical systems, 1D near-wavelength gratings have already been considered to suppress the coating thermal noise of gravitational-wave-interferometer mirrors \cite{Bruckner2008, Bunkowski2006}, to better control the spectral properties of VCSEL lasers \cite{Chang-Hasnain2011}, or for on-chip high-numerical-aperture focusing elements~\cite{Chang-Hasnain2011}.

In this work, we have used a 2D near-wavelength diffraction grating in the form of a square lattice of circular holes in an Indium Phosphide membrane (optical index of 3.3 at 1064 nm, which is the target operation wavelength). This approach is expected to preserve the main characteristics of the membrane even if most of the surface is covered with holes. Note that contrary to the 1D-case, the symmetry of the problem ensures that the grating reflectivity is independent of the incoming polarization. Figure 1b shows a color plot of the reflectivity as a function of membrane thickness and wavelength, computed by Rigorous Coupled Wave Analysis (RCWA) using the freely available software package S$^4$ \cite{Liu20122233} for a square pattern of periodicity $\Lambda=725$ nm, a hole radius $r_\mathrm{h} = 201$ nm and a material index n = 3.3. Three regions can be identified: for $\lambda > \Lambda n_\text{eff}$, with $n_\text{eff} \approx 2.9$,  a single mode propagates inside the slab. In this regime, the Fabry-Perot effect between the two sides of the slab gives rise to a periodic pattern with a reflectivity varying between 0 and $\left(\left(1-n_\text{eff}^2 \right)/ \left(1+n_\text{eff}^2\right) \right)^2 \approx 62 \%$ as a function of membrane thickness. For $\lambda<\Lambda$, optical power lost by diffraction in higher-order modes precludes the possibility of high-reflectivity. Finally, in the near-wavelength region  $\Lambda<\lambda<\Lambda n_\text{eff}$, clear signatures of the interference between 0- and higher-order modes are visible under the form of Fano line-shapes, giving rise to high-reflectivity regions. Contrary to the 1D scenario, where the physics of the problem is essentially captured by a two-mode model \cite{Karagodsky2010}, three families of non-degenerate modes have to be taken into account (see Figure 1a). The degeneracy between the (0,1) and (1,0) modes is lifted by the choice of the incoming polarization (obviously, changing the incoming polarization from TE to TM would exchange the role of these modes). The horizontal black line in Figure 1b corresponds to the particular choice we have made for the thickness (d = 265 nm). This value has been chosen because it creates a large reflectivity plateau around 1064 nm, which is expected to mitigate the influence of fabrication imperfections.

\begin{figure*}
\centering
\includegraphics[height=5.5cm]{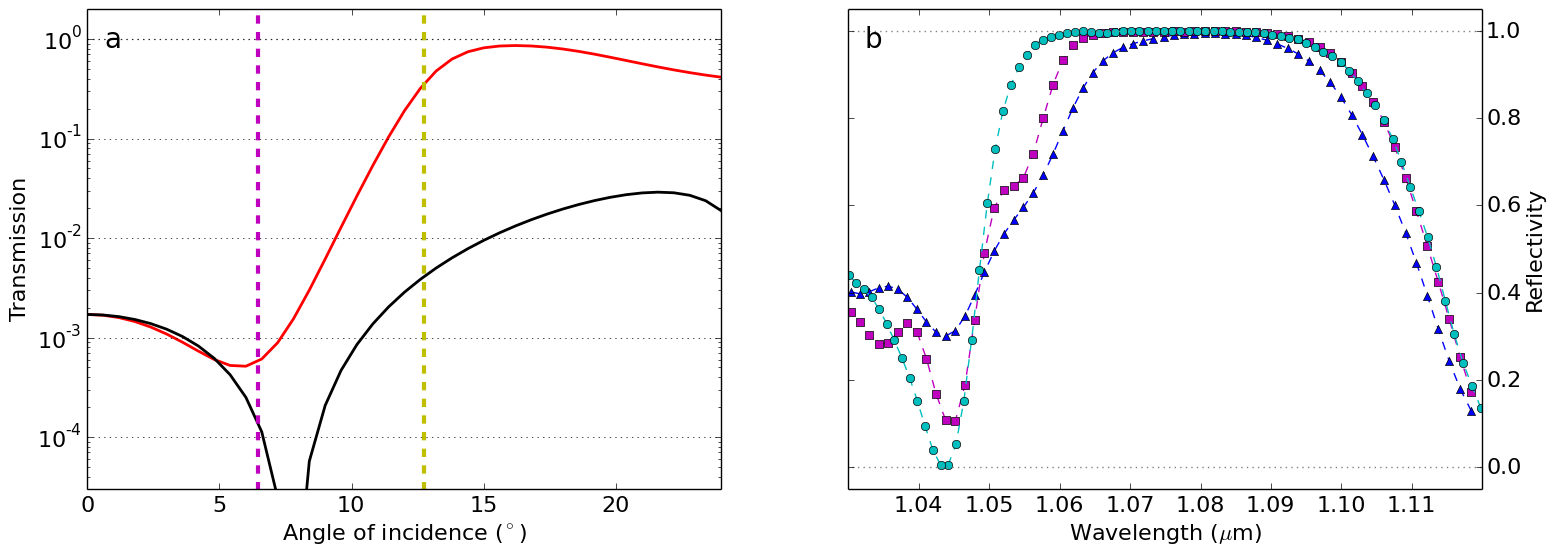}
\caption{a: Transmission of an infinite membrane as a function of the angle of incidence (red: TE-polarization; black: TM-polarization). The purple (yellow) dashed line represents the angular dispersion of a 1064 nm beam of waist $w_0 = 3\,\mu$m  ($1.5\, \mu$m) b: Reflectivity of a NWG simulated via FDTD for the parameters described in Figure 1b and $d = 265 \,\rm{nm}$, for three different configurations: infinite grating illuminated by a plane wave (green circles), and $30 \times 30 \,\mu$m grating illuminated by a beam with a waist of $3 \,\mu$m (purple squares) and $1.5 \,\mu$m (blue triangles).}
\label{FDTDinf}
%\end{center}
\end{figure*}

To reduce the effective mass of the device, the transverse dimensions of the suspended membrane are chosen as small as $30 \,\mu$m$ \times 30 \,\mu$m, yielding a mass $m \approx  100$ pg. The previous computations assume a perfect plane wave incident on an infinite grating but in the realistic scenario of a laser beam impinging on the finite membrane, one has to take into account the angular spread in the wavevector associated with the beam waist on the membrane. As an illustration, in Figure 2a, we have computed by RCWA the grating reflectivity  as a function of the angle of incidence (for completeness, we have simulated the effect for electric or magnetic fields parallel to the membrane). As a reference, the angular dispersions $\Delta \theta =\arctan(\lambda /\pi w_0)$ corresponding to a beam waist $w_0=1.5$ and 3 $\mu$m are represented on the same figure by a yellow and purple dashed line respectively. This simulation indicates that beam waists larger than 3 $\mu$m are compatible with reflectivities in the 99 \% range. To take into account the finite size of the membrane and the real transverse shape of the laser beam, we have conducted a set of simulations using the Finite Difference in Time Domain (FDTD) method using the software package MEEP \cite{OskooiRo10}. To keep reasonable computation times in the case of a finite membrane lateral extent, where translation invariance is broken, we have limited the grid resolution to only 32 points/$\mu$m. The main effect of this reduced grid resolution is to shift the predicted reflectivity plateau towards higher wavelength (the high-reflectivity region is now centered around 1080 nm). Figure 2b shows the simulated reflectivities for Gaussian beams of waists $3 \,\mu$m and $1.5 \,\mu$m. These simulations confirm that reflectivities above 99 \% can be obtained with a beam waist of 3 $\mu$m over a wavelength span larger than 20 nm.

The NWG formed by a 2D photonic crystal have been fabricated in a 265 nm-thick suspended InP membrane. More details about the fabrication, as well as the mechanical properties of the membranes can be found in \cite{Antoni2011}. The membrane reflectivity has been measured by inserting it as end-mirror of a plano-concave Fabry-Perot cavity. To avoid clipping losses on the $30\,\mu \mathrm{m} \times 30 \,\mu \mathrm{m}$ membrane, we have realized an optical cavity with a waist as small as $w_0 \approx 3 \,\mu$m. Ultra-small waists can be achieved with a small radius of curvature (RoC) coupler. These specifically designed couplers have been obtained by $\text{CO}_2$ laser photoablation on a silica plate \cite{Kuhn2014} (Herasil from HTM-Heraeus). The photoablated concavity has subsequently been coated with dielectric layers featuring a transmittance $T = 1.4$ \%  for a wavelength band of 100 nm around 1064 nm.

\begin{figure}[h!]
\includegraphics[width=7.5cm]{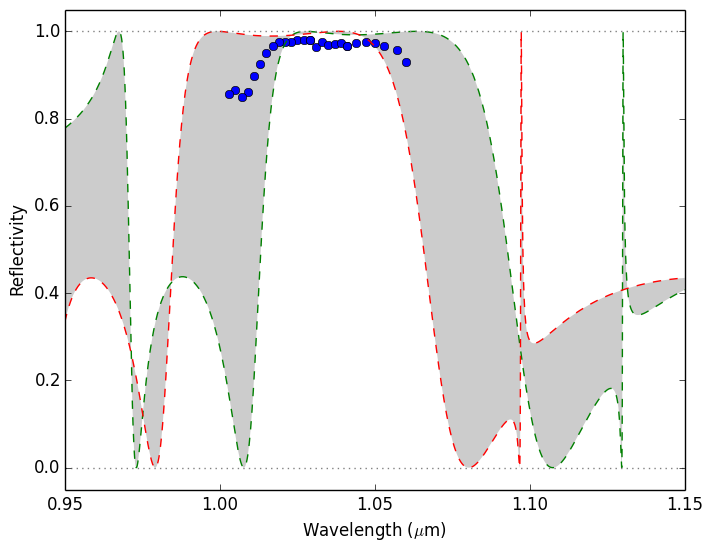}
\caption{Reflectivity inferred from a finesse measurement as a function of laser wavelength. Error bars are smaller than the points. Theoretical curves are obtained by RCWA for the same geometrical parameters as in Figure 1b, assuming a refractive index of 3.1 (dashed red) and 3.2 (dashed green). The gray-shaded area represents the corresponding uncertainty region.}
\label{cavite}
\end{figure}

The cavity is made of the membrane mounted on a 3-axis nanopositioning stage  (attocube GMBH) and set about $120 \,\mu$m away from the input coupling mirror with a 160-$\mu$m RoC placed on a two-axis mount. The mode matching between the laser beam and the cavity mode achieved with a microscope objective is close to $\eta_\text{cav}=70 \%$. We have taken advantage of one of the mechanical resonances of the Z axis of the nanopositioner to sweep the cavity length over several free spectral ranges. A Littman-Metcalf diode laser (Lion System from Sacher GMBH) is used to measure the finesse of the cavity as a function of wavelength.

Figure 3 shows the reflectivity $R$ inferred from the measurement of the finesse $\mathcal{F}=2 \pi/(T+1-R)$. A large region of high reflectivity is visible between 1020 nm and 1060 nm, as expected from the simulations. The main source of uncertainty in the experimental parameters is the refractive index of InP for which different values have been reported in the litterature \cite{Adachi1989, Bass2009}. Small changes of this parameter mainly result in a global shift of the reflectivity spectrum. To illustrate this effect, we have conducted two RCWA simulations with identical geometric parameters and different refractive index values separated by 0.1 (Figure 3). Our measurements are compatible with a refractive index close to 3.15 as in Ref. \cite{Adachi1989}.

The highest cavity finesse of $190 $ at $\lambda = 1.025 \,\mu$m corresponds to a reflectivity close to 98.1 \%, probably limited by non-periodic defects in the microfabrication process. These measurements are taken in the undercoupled regime ($T<1-R$). We have used the fact that the reflection coefficient at resonance is strongly correlated with the finesse in this regime to check that $T$ is indeed independent of wavelength over the wavelength-span of interest.

\begin{figure}
\includegraphics[width=7.5cm]{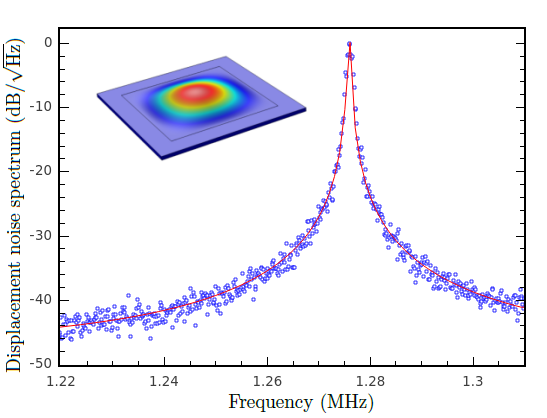}
\caption{Thermal noise spectrum of the fundamental mode of a suspended membrane with a 2D NWG grating. Inset: Corresponding color-coded strain obtained by finite-element modeling.}
\label{cavite}
\end{figure}

We have used the microcavity to probe the thermal motion of the fundamental mechanical mode of the suspended membrane (see FEM simulation in the inset of Figure 4). The cavity is locked at resonance with a lock-in amplifier and the phase fluctuations of the beam reflected by the cavity are monitored by a homodyne detector and sent to a spectrum analyzer. The Brownian noise of the suspended nanomembrane is shown in Figure 4. A Lorentzian fit of the data gives a mechanical resonance frequency $\Omega_\mathrm{m}/2 \pi = 1.27\,\rm{MHz}$ (the simulation predicts a resonance frequency close to 1.1 MHz) and mechanical quality factor of $Q \approx 500$. These Q-factors can be enhanced up to 60~000 by inducing tensile stress in the membrane.

Remarkably, the epitaxial growth used in this work is entirely compatible with the fabrication of a crystalline Bragg mirror in-situ below the membrane to form an integrated version of the membrane-in-the-middle setup. Moreover such membranes can be actuated using dielectric forces in the vicinity of an interdigitated capacitor \cite{Unterreithmeier2009, electrostatic}. Hence, our resonators could be coupled at the same time to a microwave and an optical cavity and used as transducers of (quantum) information between these fields. Such a hybrid link could play a key role in quantum information networks to interface superconducting quantum circuits with optical photons that can be transported at room temperature in optical fibers without undergoing significant decoherence.

\paragraph{Acknowledgements:}we thank R. Gu\'erout for fruitful scientific discussions. This research has been partially funded by the ``Agence Nationale de la Recherche'' program ``ANR-2011-B504-028-01 MiNoToRe'', the Marie Curie Initial Training Network cQOM, and the DIM nano-K \^Ile-de-France program ``NanoMecAtom''.

\end{document}